\documentstyle[prl,tighten,aps,epsf,amstex,multicol,times,bbm]{revtex}
\newcommand{\twovec}[2]{\left(\begin{array}{c} #1 \\ #2 \end{array}\right)}
\newcommand{\twomat}[4]{\left(\begin{array}{cc} #1 & #2 \\ #3 & #4\end{array}\right)}
\newcommand{\C}{{\Bbb C}}
\newcommand{\identity}{\openone}
\draft
\begin{document}
\author{K. Audenaert$^{1}$, J. Eisert$^{2,3}$, E. Jan\'e$^{2,4}$, M.B.
Plenio$^{2}$, S. Virmani$^{2}$ and B. De Moor$^{1}$}
\title{The asymptotic relative entropy of entanglement}
\address{
$^{1}$ Katholieke Universiteit Leuven, Dept. of Electrical Engineering
(ESAT-SISTA), B-3001 Leuven-Heverlee, Belgium\\
$^{2}$ QOLS, Blackett Laboratory, Imperial College of Science, Technology
and Medicine, London, SW7 2BW, UK\\
$^{3}$ Institut f{\"u}r Physik, Universit{\"a}t Potsdam, 14469
Potsdam, Germany\\
$^{4}$ Departament d'Estructura i Constituents de la Mat\`eria,
Universitat de Barcelona, E-08028 Barcelona, Spain}
\date{\today}
\maketitle

\begin{abstract}
We present an analytical formula for the asymptotic relative
entropy of entanglement w.r.t. PPT states for Werner states of
arbitrary dimension. We then demonstrate its validity using
methods from convex optimization. This is the first case in which
the asymptotic value of a subadditive entanglement measure has
been calculated.
\end{abstract}


\begin{multicols}{2}
The systematic investigation of quantum entanglement is a major
goal of quantum information theory \cite{Plenio V 98}. In the
study of entanglement manipulation one considers the
interconversion of different forms of entanglement by means of
local quantum operations and classical communication (LQCC). For
pure bi-partite states entanglement manipulation in both the
finite and asymptotic limits is quite well understood. For pure
states necessary and sufficient conditions for the local
interconvertibility of entangled states are known. In the
asymptotic limit of infinitely many copies of a pure state, a
single number, the von Neumann entropy of a subsystem,
appropriately quantifies the degree of entanglement
\cite{purestates}.

Much less is known about the entanglement of mixed states.
 One approach is to define {\it entanglement
measures}, which are functions of a state that cannot increase
under local operations and provide constraints on possible local
entanglement manipulation protocols. These measures prove to be
useful mathematical and conceptual tools, and have interesting
links with other areas such as the study of channel capacities
\cite{entform}.  A number of such measures have been proposed,
most notably the entanglement of formation \cite{entform}, the
distillable entanglement \cite{entform,Rains 99}, and the relative
entropy of entanglement \cite{Rains 99,Vedral PRK 97}. The
distillable entanglement is defined as the asymptotic number of
pure maximally entangled states that can be obtained via LQCC from
a supply of a given state. For mixed states the distillable
entanglement is exceedingly difficult to compute as it is defined
as an asymptotic quantity referring to infinitely many copies of a
quantum state. Therefore, upper bounds on the distillable
entanglement, in particular other entanglement measures, are of
major practical use. One such entanglement measure is the relative
entropy of entanglement, defined as
\begin{equation}\label{Rele}
        E_R(\sigma)=\min_{\rho\in{\cal D}} S(\sigma||\rho),
\end{equation}
for states $\sigma$, where ${\cal D}$ is the
set of states with positive partial transpose %\cite{PPT}
(PPT states), and $S(\sigma||\rho)=\text{tr}[\sigma\lg \sigma-
\sigma\lg \rho]$ ($\lg$ signifies $\log_2$).  This function
essentially quantifies the distinguishability of $\sigma$ from the
set of PPT states. The set ${\cal D}$ can also be taken to be the
set of separable states \cite{Vedral PRK 97}. However, the set of
PPT states is much easier to characterise, and the resulting
measure provides a tighter bound to the distillable entanglement,
one that is actually attained on pure states and certain mixed
states \cite{entform}.

In general, efficient protocols for entanglement
manipulation require an asymptotic number of copies of the initial state.
Therefore, to address any question related to asymptotic entanglement manipulation,
 one will instead have to consider asymptotic versions of the entanglement measures.
For a given measure of entanglement $E$, the asymptotic version $E^{\infty}$
is defined as the average entanglement per copy in the asymptotic limit \cite{Limit,Hayden 01},
\begin{equation}
    E^{\infty}(\sigma) = \lim_{n\rightarrow \infty}
    \frac{E(\sigma^{\otimes n})}{n}.
\end{equation}
For example, the asymptotic cost of creating a mixed state by LQCC
from a supply of pure maximally entangled states
is given by the asymptotic entanglement of formation \cite{Hayden 01}.
However, such asymptotic entanglement measures are difficult to compute, and so far
this task has not been accomplished except for the very rare occasions where
the measure in question is known to be additive\cite{Add}.
\begin{figure}[hbt]
\begin{center}
    \leavevmode
    \epsfysize=4.2cm
    \epsfbox{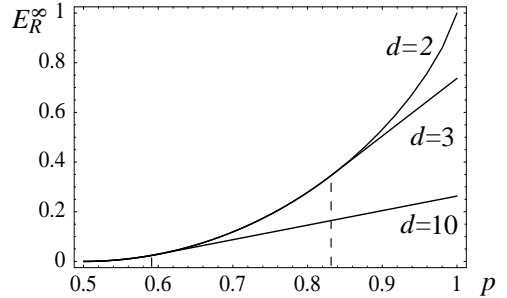}
    \caption{The asymptotic relative
entropy of entanglement $E_R^\infty(\sigma(p))$
of Werner states in $\C^d\otimes \C^d$
as a function of  $p$  (the weight of the antisymmetric state) for
several values of $d$. The dashed lines separate
the two regimes of Eq.\ (\ref{theorem}), for $d=10$ and for $d=3$.
}
    \label{fig1}
\end{center}
\end{figure}
In this Letter we present a general formula
for the asymptotic relative entropy
of entanglement for an important class
of bi-partite states, namely the Werner states of
arbitrary dimension \cite{Werner89}. These states are the only
states that are invariant under local
unitaries of the form $U\otimes U$.
They can be conveniently formulated in terms
of the projectors $\pi_0$ ($\pi_1$)
on the symmetric (antisymmetric) subspaces of a Hilbert space
${\cal H}=\C^d\otimes \C^d$.
Denoting the permutation operator that interchanges the subsystems as
$\pi$, the projectors can be expressed as
$\pi_0=(\identity+\pi)/2$ and $\pi_1=(\identity-\pi)/2$.
A general Werner state is of the form
$\sigma(p) = p\sigma_1 + (1-p)\sigma_0$,
$p\in[0,1]$, where
$\sigma_0=\pi_0/\text{tr}[\pi_0]$ and
$\sigma_1=\pi_1/\text{tr}[\pi_1]$.
Appropriately exploiting the
symmetry of this state
is one of the key ingredients to the proof of
the main statement of this Letter.
We formulate this main result as a theorem.\\
{\bf Theorem.}
For states in $\C^d\otimes \C^d$
of the form $\sigma(p) = p\sigma_1 +
(1-p)\sigma_0$ with $p\in (1/2,1]$,
$E_R^\infty$ w.r.t. PPT states is given by
%
%For any state of the form $\sigma(p) = p\sigma_1 +
%(1-p)\sigma_0$ of a bipartite system with $d$-dimensional
%subsystems the $E_R^\infty$ is
%given by
\begin{eqnarray}
    E^{\infty}_R(\sigma(p)) =
        \left\{ \begin{array}{ll}
    1-H(p), & p\leq \frac{d+2}{2d} \\
        & \\
    \lg\frac{d+2}{d}+(1-p)\lg\frac{d-2}{d+2}, & p>\frac{d+2}{2d}
    \end{array} \right.\label{theorem}
\end{eqnarray}
where $H(p) = -p\lg p-(1-p)\lg(1-p)$. \\

It is interesting to note from this formula that $E_R^\infty$ never
exceeds the logarithm of the negativity \cite{logneg}.
% and hence this becomes the tightest known upper bound on
%the distillable entanglement of the Werner states.
Curiously, $E_R^\infty(\sigma(p))$ is a convex function of $p$,
whereas numerics show that $E_R(\sigma(p)^{\otimes n})/n$ is not
convex for any finite number $n>1$. It is also intriguing that
this formula is {\it exactly} the same as that from a different
optimization problem considered by Rains \cite{semidef}, where he
minimizes the function $B(\rho,
\sigma)=S(\rho||\sigma)+\log|\sigma^{\Gamma}|$ over {\it all}
states $\sigma$. Rains' quantity is also an upper bound to the
distillable entanglement. However, it is not a convex function,
which makes its minimization difficult due to the possibility of
local minima. This applies especially in the asymptotic limit as
the additivity properties of $B(\rho,\sigma)$ are unknown.

{\bf Proof:} The proof proceeds in two stages. First
we provide an upper bound on
$E_R(\sigma^{\otimes n})/n$ for any number of copies $n$, by presenting a trial optimal state.
Then we derive
a lower bound for
$E_R(\sigma^{\otimes n})/n$ using convex optimization methods.
We conclude by showing that the
two bounds coincide in the asymptotic limit
$n\rightarrow \infty$.

{\em Upper bound:}
We consider a situation where we hold $n$ copies of a Werner state
$\sigma(p)$. Following the work of Ref.\ \cite{WernerSymm}, we will make
heavy use of symmetry.  In
Refs. \cite{Rains 99,WernerSymm} it has been shown that if a state is
invariant under a certain symmetry group, then one can restrict
the minimization in Eq.\ (\ref{Rele}) to those PPT states that
are also invariant under the same group.
As $\sigma(p)$
is invariant under the group $G$ of local unitaries
of the form $U\otimes U$, the state $\sigma(p)^{\otimes n}$
is invariant under $G^{\otimes n}$.
This implies that the optimal PPT state for
$\sigma(p)^{\otimes n}$ can be chosen to be
a convex sum of the $2^{n}$ possible
$n$-fold tensor products of $\sigma_{0}$ and $\sigma_{1}$,
\begin{equation}\label{form}
        \eta =
\sum_{f=0}^{2^{n}-1} x_f
(\sigma_{f_{1}}\otimes \sigma_{f_{2}}\otimes \cdots \otimes \sigma_{f_{n}}),
\end{equation}
where $\vec{x}=(x_0,...,x_{2^{n}-1})^T$ forms a probability distribution,
and $f_{i} \in \{0,1\}$, $i=1,...,n$,
is the $i$-th digit in
the binary representation of $f$.
Furthermore, as the state $\sigma^{\otimes n}$ is
invariant under any permutation of the $n$ copies, we can
add the constraint that $x_{l}=x_{m}$ if the number of 1's in the
binary representations of $l$ and $m$ are the same.

We will consider trial states of the form
$\sum_{i} w_{i} (u_{i} \sigma_{1} + v_{i} \sigma_{0})^{\otimes n}$
, which automatically satisfy this constraint. Such states correspond to:
\begin{equation} \label{ecs}
    \vec{x}=\sum_i w_i \twovec{u_i}{v_i}^{\otimes n}
\end{equation}
where the $u_i(v_i)$ component is the weight of
$\sigma_1(\sigma_0)$, $u_i+v_i=1$ and $\sum_i w_i =1$ (and
therefore $\sum_j x_j = 1$).

We will need to know the eigenvalues of the partial transpose of $\eta$ in
order to ensure that it is PPT. Ignoring degeneracy, there are only two
eigenvalues of the partial transpose of the Werner state $\sigma(p)$. It
is easy to show that they are non-negative iff the following two component
vector is non-negative:
\begin{equation}
  \twomat{ -1 }{ 1}{ 1}{(d-1)/(d+1)} \twovec{p}{1-p}=  T\twovec{p}{1-p}.
\end{equation}

Similarly, it can easily be shown that $\eta$ will be a PPT state iff the $\vec{y}$ and
$\vec{x}$ in the following equation are non-negative vectors:
\begin{equation} \label{cond}
    \vec{y} = \twomat{ -1 }{ 1}{ 1}{(d-1)/(d+1)}^{\otimes n}
    \vec{x} = T^{\otimes n} \vec{x}
\end{equation}
Subject to this condition we need to calculate the relative entropy
between the $n$-copy Werner state $\sigma(p)^{\otimes n}$ and this
generalized Werner state $\eta$. This is given by:
\begin{eqnarray}
    S&&(\sigma(p)^{\otimes n}||\eta)/n\nonumber\\
    &&=- H(p)
    - (1/n)\sum_{k=0}^n C^k_n p^{n-k} (1-p)^k \lg\sum_i w_i u_i^{n-k}
    v_i^{k}.\label{relent1}
\end{eqnarray}

At this point, we notice that since the second term in
Eq.\ (\ref{relent1}) is the average of the function
\begin{equation} \label{xi}
\xi(n-k) = (1/n)\lg\sum_i w_i u_i^{n-k} v_i^{k}
\end{equation}
over a binomially distributed
variable $k$, the value can be substituted by $\xi(np)$ when we
take the limit $n\rightarrow\infty$ \cite{rigormortis}.

Since $E^\infty_R(\sigma(p)) $ is the minimal value of
$S(\sigma(p)^{\otimes n}||\eta)/n$ over all possible PPT states
$\eta$, any such PPT state $\eta$ will give us an upper bound for
$E_R^\infty(\sigma(p))$.
In particular, for the vector
$\vec{x}$ we propose a mixture of two $n$-fold Kronecker
powers:
\begin{equation}
    \vec{x} = \sum_{i=1}^{2} w_i\twovec{1-a_i}{a_i}^{\otimes n},
\end{equation}
where $0\le w_i\le 1$, $w_1=1-w_2$, $0\le a_1\le 1/2$ and $1/2\le
a_2\le 1$. Proper values for the parameters need to be selected,
to ensure that the corresponding state $\eta$ will be PPT. Inspired by
numerical results, we consider two separate intervals for
$p$ in Eq.\ (\ref{theorem}); $1/2\le p\le
(d+2)/(2d)$, and $(d+2)/(2d)\le p\le 1$.

For the first interval, set $w_{1}=0$ and $a_2=1/2$, so that
$x_k=2^{-n}$ for all $k$.
 This state gives us an upper bound that equals $E_{R}$ for one
copy of $\sigma(p)$
\begin{equation}\label{onecopy}
    E_R^{\infty}(\sigma(p)) \le \frac{S(\sigma(p)^{\otimes n}||\eta)}{n}
    = 1-H(p).
\end{equation}

Now consider the second interval. We will set
\begin{eqnarray}\label{choice}
    a_1 &=& \frac{(d+2)(1-p)}{d+2-4p}\; ,\;a_2 =
\frac{1+d-(d+2)a_1}{d+2-4a_1}\nonumber\\
    w_1 &=& 1-w_2 = \frac{1}{1+z^n} \; ,\;z = (d+2-4a_1)/d.
\end{eqnarray}
We calculate $T^{\otimes
n}\vec{x}$:
\begin{eqnarray}
    \vec{y}=T^{\otimes n}\vec{x} &=& \sum_{i=1}^2
    w_i \twovec{2a_i-1}{1-2a_i/(d+1)}^{\otimes n}
\end{eqnarray}
so that, with the values of Eq.\ (\ref{choice})
\begin{equation}
    y_k = \sum_{i=1}^{2}
w_i(2a_i-1)^{n-\#(k)}\left(1-\frac{2a_i}{d+1}\right)^{\#(k)},
\end{equation}
where $\#(k)$ is the number of 1's in the binary representation of $k$.
It is easy to check using Eq.\ (\ref{choice}) that $y_k$ is always
non-negative.

As a consequence, by taking into account the discussion after
Eq.\ (\ref{relent1}) and the notation introduced in Eq.\ (\ref{ecs}), the
upper bound for $n\rightarrow\infty$ reads
\begin{eqnarray}
    E_{R}^{\infty}(\sigma(p)) &\le& -H(p) - \lim_{n\rightarrow\infty}
    \lg \left(\sum_{i=1}^2 w_i u_i^{pn} v_i^{(1-p)n}\right)^{1/n}\nonumber
\\
    &=& -H(p) - \lg\lim_{n\rightarrow\infty}
\left(\frac{t_1^n+t_2^n}{q_1^n+q_2^n}
    \right)^{1/n}\nonumber\\
    &=& - H(p) - \lg\frac{\max(t_1,t_2)}{\max(q_1,q_2)} \label{max}
\end{eqnarray}
with
\begin{eqnarray}
    t_1 &=& d(d-2)^p (d+2)^{1-p} (1-p)^{1-p} p^p \nonumber\\
    t_2 &=& (d-2+d^2(1-p))^p (d^2 p-d-2)^{1-p} \nonumber\\
    q_1 &=& d^2-4 \;\; \mbox{and} \;\; q_2 = d(d+2-4p).
\end{eqnarray}
It is easy to check that for $p\geq(d+2)/(2d)$ (the second
interval), both $t_1\geq t_2$ and $q_1\geq q_2$. So we obtain:
\begin{equation} \label{upper2}
    E_{R}^{\infty}(\sigma(p)) \le -H(p) - \lg\frac{t_1}{q_1}
    = \lg\frac{d-2}{d} + p \lg\frac{d+2}{d-2}.
\end{equation}
In the other regime, $p\leq(d+2)/(2d)$, the bound obtainable from Eq.\ (\ref{max})
is worse than Eq.\ (\ref{onecopy}). This
ends the proof of the upper bound.

{\em Proof of lower bound:} We now proceed to find a lower bound on
$E_R$. To do this, we will use the idea of Lagrange duality
\cite{Boyd V 00}. To calculate $E_R^\infty$ we need to solve
the optimization problem
\begin{eqnarray}
   &&\frac{ E_{R}(\sigma(p)^{\otimes n})}{n} = \min_{\vec{x}}\{ -H(p)-
    \frac{1}{n}\vec{z}^T\cdot\lg \vec{x}\} \nonumber\\
&& \mbox{with} ~~~~~  \vec{x} \ge 0 \hspace{5pt},\hspace{10pt} -T^{\otimes
n}\vec{x} \le0 \hspace{5pt},\hspace{10pt}
    \sum_{k=0}^{2^n-1} x_k=1,
\end{eqnarray}
where $\vec{z}^T = (p,1-p)^{\otimes n}$. This is a
convex optimization problem, so it is possible to consider the
so-called dual problem.
It is a basic result in convex optimization theory that the
optimal (maximal) $g$-value of the dual problem is a lower bound
on the optimal (minimal) value of the primal problem, which is just what we are
looking for (see \cite{Boyd V 00}
for a general description of duality in optimization).

The dual problem can be obtained as follows.
First form the Lagrangian by multiplying the constraints with
Lagrange multipliers and adding them to the objective function:
\begin{eqnarray}
    {\cal L}(\vec{x},\vec{\lambda},\nu) &=& -H(p) -
    \frac{1}{n}\vec{z}^T \lg \vec{x} \nonumber \\
    &&-\vec{\lambda}^T  T^{\otimes n}\vec{x} + \nu(\sum_{k=0}^{2^{n}-1}x_k - 1).
\end{eqnarray}
The constraint $\vec{x} \ge 0$ is not included explicitly, as
it just determines the domain of the function $\lg$. Note that
$\vec{\lambda}$ must exhibit the same copy-permutation symmetry as
 $\vec{x}$. Because the
constraint associated to $\vec{\lambda}$ is an inequality, we
have to introduce the constraint $\vec{\lambda}\ge 0$.
The dual function is now given by
\begin{equation}
    g(\vec{\lambda},\nu) = \inf_{\vec{x}\ge 0} {\cal
L}(\vec{x},\vec{\lambda},\nu),
\end{equation}
and the dual optimization problem is:
\begin{equation}
    \max_{\vec{\lambda},\nu} g(\vec{\lambda},\nu)\hspace{5pt},
\hspace{10pt} \vec{\lambda} \ge 0,
\end{equation}
including any other constraints on the domain of $g$.

For our problem, the dual function can be calculated
explicitly. The derivative of the Lagrangian w.r.t.\ $x_k$ is
\begin{equation}
    \frac{\partial {\cal L}}{\partial x_k} = -\frac{1}{n \ln
    2}\frac{z_k}{x_k} - \left(\vec{\lambda}^T\cdot
    T^{\otimes n}\right)_{k}\; +\nu.
\end{equation}
The Lagrangian reaches an extremum (minimum) at
\begin{equation}
    \hat{x}_k =  \frac{1}{\mu_k}\frac{z_k}{n \ln 2},
\end{equation}
where $\mu_k=\nu- \left(T^{\otimes
n}\vec{\lambda}\right)_{k}$ and we have exploited the symmetry of
$T$. The dual function is
\begin{eqnarray}
 g(\vec{\lambda},\nu) &=&\frac{1+\ln(n\ln2)}{n \ln2}
+\frac{1}{n}\sum_{k=0}^{2^{n}-1}z_k\lg\mu_k-\nu
\end{eqnarray}
where we have used that $\vec{z}^T = (p,1-p)^{\otimes n}$, which
implies that  $ \sum_{k=0}^{2^{n}-1} z_k \lg z_k = -nH(p)$ . As
stated before, $\vec{\lambda}$ must be non-negative, and
inspecting the domain of $g(\vec{\lambda},\nu)$ yields an
additional constraint that the $\mu_{k}$ be non-negative.

Now, any acceptable assignment of values to $\vec{\lambda}$ and $\nu$
gives a lower bound to $E_R^\infty(\sigma(p))$.
Again we consider the two $p$-intervals of Eq.\ (\ref{theorem}).
For $  p \le  (d+2)/(2d)$ we propose
\begin{eqnarray} \label{point1}
    \nu^1 &=& 1/(n\ln2) \nonumber\\ \vec{\lambda}^1 &=&
    \frac{\nu^{1}}{d^n}\left( \twovec{1}{d+1}^{\otimes n} -
    \twovec{d+1-2dp}{d+1}^{\otimes n} \right).
\end{eqnarray}
After a short calculation we obtain $ \mu_k = \nu^{1} 2^n
p^{n-\#(k)}(1-p)^\#(k)$ . This gives a feasible point of the dual
problem, because both $\vec{\lambda}$ and $\mu_{k}$ as given here are
non-negative. The value of $g$ using these assignments is
%\begin{equation} \label{low1}
 $  g(\vec{\lambda}^1,\nu^1) = 1 - H(p).$
%\end{equation}

For the second interval, $p\geq p'=(d+2)/(2d)$, we replace $p$ by $p'$
in Eq.\ (\ref{point1}), giving
\begin{eqnarray}
    \nu^2 =\nu^1\,,\, \vec{\lambda}^2 =
    \frac{\nu^{2}}{d^n}\left( \twovec{1}{d+1}^{\otimes n} -
    \twovec{-1}{d+1}^{\otimes n} \right).
\end{eqnarray}
We now obtain
%\begin{equation} \label{low2}
 $   g(\vec{\lambda}^2,\nu^2) =
 \lg\frac{d+2}{d}+(1-p)\lg\frac{d-2}{d+2}.$
%\end{equation}

As the two lower bounds $g(\vec{\lambda}^1,\nu^1)$ and
$g(\vec{\lambda}^2,\nu^2)$ coincide with the two upper bounds
(Eqs.\ (\ref{onecopy}) and (\ref{upper2})), the proof of the
Theorem is now complete.\hfill\fbox{}

The remarkable behavior of $E_{R}^{\infty}(\sigma(p))$ is shown in
Fig.\ \ref{fig1} for several values of $d$. The nonlinear behavior
for small values of $p$ goes over into a linear dependence on $p$
above the threshold value $p'=(d+2)/2d$. An immediate consequence
of the result is that there are no inseparable Werner states with
zero entanglement cost---a similar conclusion could not be drawn
from Rains' bound as it is not an asymptotic quantity. It is
astonishing that, as long as $p\le p'$, $E_R^\infty$ is invariant
under the strongly irreversible operation of twirling, mapping
Werner states on $\C^2\otimes \C^2$ to Werner states on
$\C^d\otimes \C^d$  \cite{Twirling}.

Interestingly, the dependence of $E_R^\infty(\sigma(p))$
on $p$ is quite similar to the conjectured
behavior of
the entanglement of formation for a  single
copy of an isotropic state \cite{Vollbrecht}:
there, one can also distinguish between
two regimes, and
for larger values of the weight $F$
of the maximally
entangled state in the isotropic state the
dependence of the entanglement of formation is
conjectured to be linearly dependent on $F$.

In this Letter we have concentrated on
the important class of Werner states. With similar
methods, one can also investigate other classes with high symmetry. It is
hoped that this work can significantly contribute
to the quest for a better understanding of the
asymptotic regime of entanglement.

We would like to thank R.F. Werner and D. Jonathan for fruitful
discussions, and E. Rains for drawing to our attention possible connections with \cite{semidef}.
 This work was partially supported by EPSRC, The Leverhulme
Trust, DFG, the European Union EQUIP project\& the ESF QIT program
and grants MEC (AP99), IUAP-P4-02 and GOA-Mefisto-666 .

% EJ acknowledges a grant by MEC (AP99). KA
%and BDM acknowledge funding from grants IUAP-P4-02 and GOA-Mefisto-666.

\end{multicols}

\begin{references}
\bibitem{Add}
If $E(\sigma^{\otimes n})=n E(\sigma)$ for all
$n=1,2,...\,$ for a given state $\sigma$, then clearly
$E^\infty(\sigma)=E(\sigma)$ holds.
\bibitem{rigormortis}
A rigorous argument is as follows. Denote the maximum absolute
value of $\partial \xi (k)/\partial k$ over $(0,n)$ by $D$. Then
the error
\begin{eqnarray*}
\epsilon&=&\left| \sum_{k=0}^n C_n^k p^{n-k} (1-p)^k \xi(n-k) - \xi(pn)
\right| \\
&\le& D \sum_{k=0}^n C_n^k p^{n-k} (1-p)^k |n-k-pn|,
\end{eqnarray*}
 because $\xi(n-k)$ lies between the bounds
$\xi(pn)\pm D|n-k-pn|$. The last term is proportional to the mean
absolute deviation from the mean, which is always less than the
standard deviation (this follows from the concavity of
$f(b)=|\sqrt{b}|$). Hence $\epsilon \leq D\sqrt{np(1-p)}$.
%For large $n$, such that the binomial distribution is sufficiently well approximated by the normal
%distribution:
%\begin{eqnarray*}
%\epsilon&\le&\int_{-\infty}^{\infty} \frac{1}{\sqrt{2\pi}\sigma}\exp(-\frac{(n-k-pn)^2}{2\sigma^2})
% D|n-k-pn| dk \\
%&=& D\sqrt{2np(1-p)/\pi}.
%\end{eqnarray*}
In the case at hand, $D$ is of order $1/n$:
from the definition of $\xi(k)$,
from Eq.\ (\ref{xi})
the maximal absolute value of $\partial \xi(k)/\partial k$ equals
$\max_i \left|\lg(u_i/v_i)\right|/n$,
and, as can be seen from Eq.\ (\ref{choice}), the values of $u_i$ and $v_i$ are independent of $n$.
Hence, in the limit $n\rightarrow\infty$, the error $\epsilon$ goes to zero as $1/\sqrt{n}$.
\bibitem{Twirling}
The twirling operation $\Pi$ in
 $d\times d$ dimensions,
$d>2$, is defined as $\Pi(\rho)=\int
d\mu_U (U\otimes U) \rho (U\otimes U)^\dagger$
for states $\rho$,
where the integral is performed with
respect to the Haar measure, $U\in U(d)$.
Under twirling,
Werner states on $\C^2\otimes \C^2$ with
weight $p$ of the antisymmetric state -- regarded
as states on $\C^d\otimes \C^d$ -- will
be mapped on Werner states on $\C^d\otimes \C^d$
with the same weight $p$.
\end{references}
\end{document}